\documentstyle[11pt,moriond,epsfig]{article}

\def\Journal#1#2#3#4{{#1} {\bf #2}, #3 (#4)}

\def\be{\begin{equation}}
\def\ee{\end{equation}}
\def\bea{\begin{eqnarray}}
\def\eea{\end{eqnarray}}

\begin{document}
\rightline{To appear in \it{Fundamental Parameters in Cosmology}\hspace{0.8cm}}
\rightline{The proceedings of the XXXIIIrd Rencontres de Moriond}
\rightline{eds.\ J. Tr\^an Thanh V\^an \& Y. Giraud-Heraud\hspace{1.98cm}}
\rightline{(Paris, France: Editions Frontieres)\hspace{3.75cm}}
\vspace*{3.5cm}
\title{GLOBAL COSMOLOGICAL PARAMETERS\\
MEASURED USING CLASSICAL DOUBLE RADIO SOURCES}

\author{RUTH A. DALY, ERICK J. GUERRA, \& LIN WAN }

\address{Department of Physics, Princeton University,\\ Princeton
NJ 08544, USA}

\maketitle\abstracts{Fourteen classical double radio galaxies with redshifts
between zero and two were used to determine the 
cosmological parameters
$\Omega_m$, $\Omega_{\Lambda}$, and $\Omega_k$, where
these are the normalized values of the mean mass density,
cosmological constant, and space curvature at the present
epoch.  A low value of $\Omega_m$ is obtained, and 
$\Omega_m = 1$ is ruled out with 97.5 \%
confidence.  The low value of $\Omega_m$ determined using
the radio source method described here is also indicated 
by several independent
tests.  Thus, it appears that either a cosmological
constant, or space
curvature, is significant at the present epoch.  
This means that the universe is undergoing, or has recently 
undergone, a transition away from a state of matter
domination and into a state where either a cosmological
constant or space curvature is determining the expansion 
rate of the universe.  The low value of $\Omega_m$ presented here 
and by Guerra \& Daly (1998) means that 
we can state with 97.5 \% confidence
that the universe will continue to expand forever.  }

\section{Introduction} 

The structure and fate of the universe can be described by the
cosmological parameters $\Omega_m$, $\Lambda$, and $k$, where
$\Omega_m$ is the average density of matter in the 
universe at the present time divided by the critical density, 
$\Lambda$ is the current value of the cosmological constant
(generally, though not always, taken to be time-independent), and
$k$ describes the global geometry of the universe.  The
normalized values of $\Lambda$ and $k$ are denoted
$\Omega_{\Lambda}$, and $\Omega_k$ (e.g. Peebles 1993).  Assuming that any
relativistic component, such as the microwave background
radiation, has a negligible contribution to the total
mass-energy density at the present time, the following 
equation must be satisfied: 
$\Omega_m$ + $\Omega_{\Lambda}$ + $\Omega_k$ = 1.  

As originally discussed by Dicke (1970), and later by 
Peebles (1993), 
each term evolves differently with redshift, so 
it is unlikely that two terms will be comparable at any given time.
However, if 
$\Omega_m < 1$, then, the universe is undergoing a transition
away from a state of matter domination into a state 
where either space curvature or the cosmological constant is
dominant.  Following Dicke's argument, it 
is rather unlikely that all three terms will be 
significant at the present time.  

The cosmological parameters can only be believably determined
when several independent methods of estimating the parameters 
all yield similar values, both within each category of test, 
and comparing results from different categories.  
At present, there are three main
categories of estimates of $\Omega_m$: (1) local, low-redshift
dynamical tests; (2) tests that depend on the coordinate distance
to high-redshift sources through the angular size distance or
the luminosity distance; and (3) tests using the fluctuations
of the microwave background radiation on different angular
scales.  At the present time, values of $\Omega_m$ have been
determined using methods (1) and (2); results obtained using
method (1) are mentioned below.  Method (2) has been applied
to radio galaxies with redshifts between zero and two
by Daly (1994, 1995), Guerra \& Daly (1996, 1998), 
and is applied here 
to radio sources with redshifts between zero 
and two.  Method (2) has also been applied to supernovae 
by Garnavich {\it et al.} (1998) and Perlmutter {\it et al.} (1998), who study 
sources to a redshift of about one.  
At some point in the future, there may be independent 
constraints from measurements of fluctuations
of the microwave background radiation on different angular scales.

Several local, low-redshift estimates of $\Omega_m$ indicate that 
it is significantly less than unity (Hudson {\it et al.} 1995; 
Shaya, Peebles, \& Tully 1995; Carlberg {\it et al.}\ 1997;
Bahcall \& Fan 1998).   
These tests are interesting
and important, though they are all subject to caveats.  
Many depend on whether the mass is distributed like the light, known as 
``biasing," and many local measurements only indicate the amount
of mass that is clustered on the scales of galaxies and clusters
of galaxies.  The concordance of many local,  
low-redshift tests suggests that the amount of mass that clusters
with galaxies is significantly less than the critical value.  

Estimates of cosmological parameters that utilize the coordinate
distance to sources at high redshift (category [2] defined above), 
such as tests involving
angular size distance or luminosity distance, 
are fundamentally different from local, 
low-redshift tests.  The value of $\Omega_m$ estimated through
the coordinate distance, the angular size distance, or the
luminosity distance, is truly the global value of $\Omega_m$.
The test is independent of any biasing of matter relative to
light, of how or whether the mass
is clustered, of the nature of the dark 
matter (e.g. baryonic, cold dark matter, hot dark matter, etc.), and of
the origin of fluctuations that lead to galaxy formation
(e.g. cosmic strings, textures, adiabatic or isocurvature
fluctuations, etc.).  

Two tests currently being used to determine global cosmological 
parameters through the coordinate distance to high-redshift
sources (category [2] defined above) 
are the radio source method, and the supernova
method.  

\section{The Radio Source Method}

A new way in which 
powerful classical double radio sources can 
be used to determine global cosmological parameters 
is described in detail
by Daly (1994, 1995), and Guerra \& Daly (1996, 1998).  
The idea is to use two independent measures of the
average size of a given radio source, where the size
is measured by the separation of the two radio hot spots.  
The two measures depend on the angular size distance to
the source in different ways, so equating them 
allows a determination of the coordinate
distance to the source, which in turn can be used to determine
values for cosmological
parameters.  

Powerful classical radio sources form an unusually homogeneous
population, and the average size, or separation
between the two radio hot spots, of the sources at any
given redshift exhibits a rather small dispersion.  One
measure of the average size of a given source 
is $<D>$, the average size of 
similar sources at the same redshift.  Another measure
of the average size of a source is $D_* =v_L~t_*$, 
where $v_L$ is the average rate of growth of that source,
and $t_*$ is the total time for which the highly collimated
outflows of that source are powered by the AGN; this outflow leads 
to the large scale radio emission.  Several observational properties
of powerful classical double radio sources can be explained
by setting $t_* \propto L_j^{-\beta/3}$ (Daly 1994).  Given this
parameterization, it is straight-forward to show that
$D_* \propto (B_La_L)^{-2\beta/3}v_L^{(1-\beta/3)}$, where
$B_L$ and $a_L$ are the magnetic field strength and width
of the radio lobe, located just behind the radio hot spot.  
Each of these parameters, $B_L, ~a_L$, and $v_L$ can 
be determined using radio data (e.g. Wellman, Daly, \& Wan 1997).  
The 
best fit value of $\beta$ is about 2, in which case
$D_* \propto (B_La_L)^{-4/3}v_L^{1/3}$, as discussed
in detail by Guerra \& Daly (1998).  

\section{Results}

\begin{figure}
\psfig{file=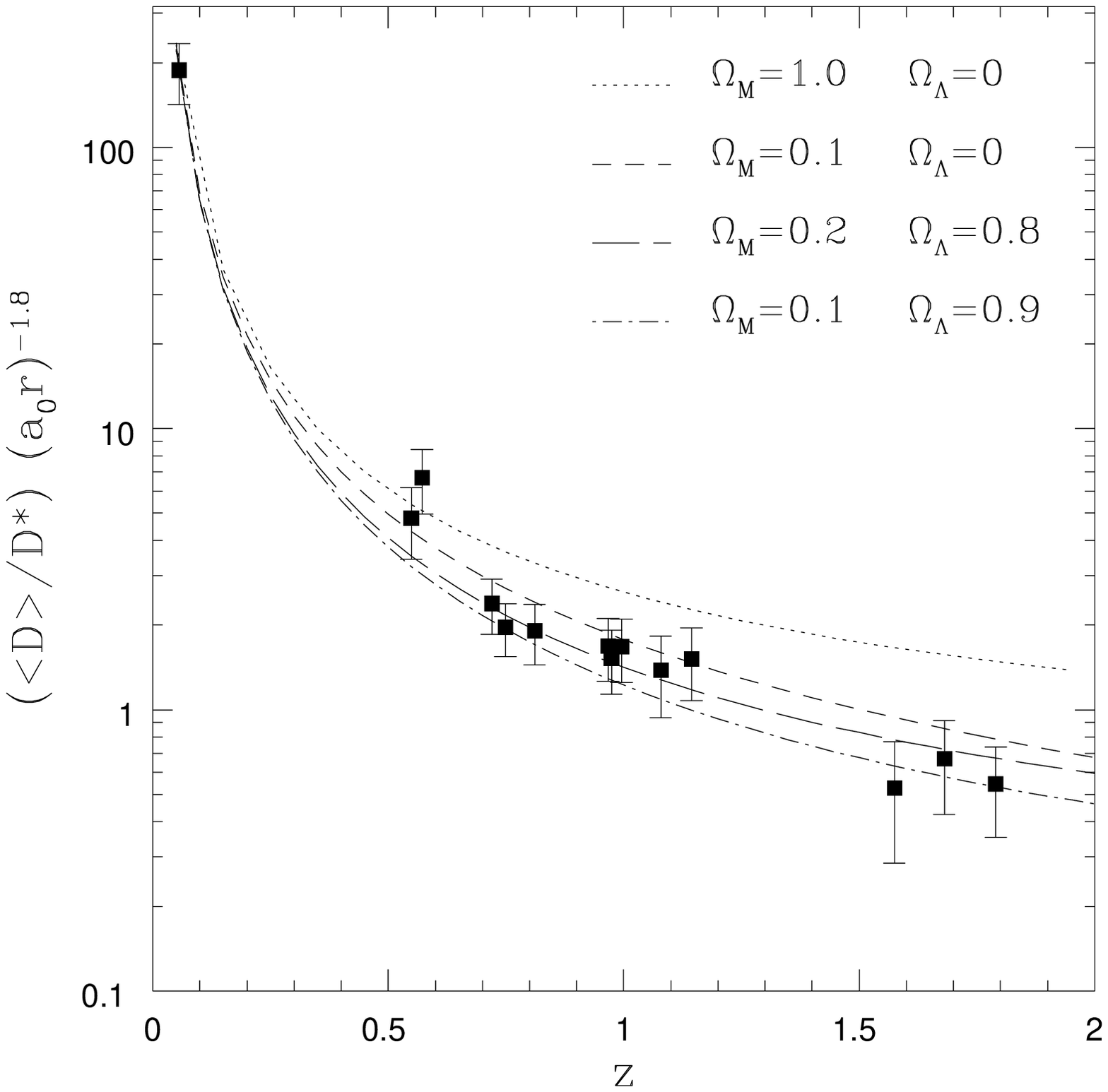,width=3.6in}
\psfig{file=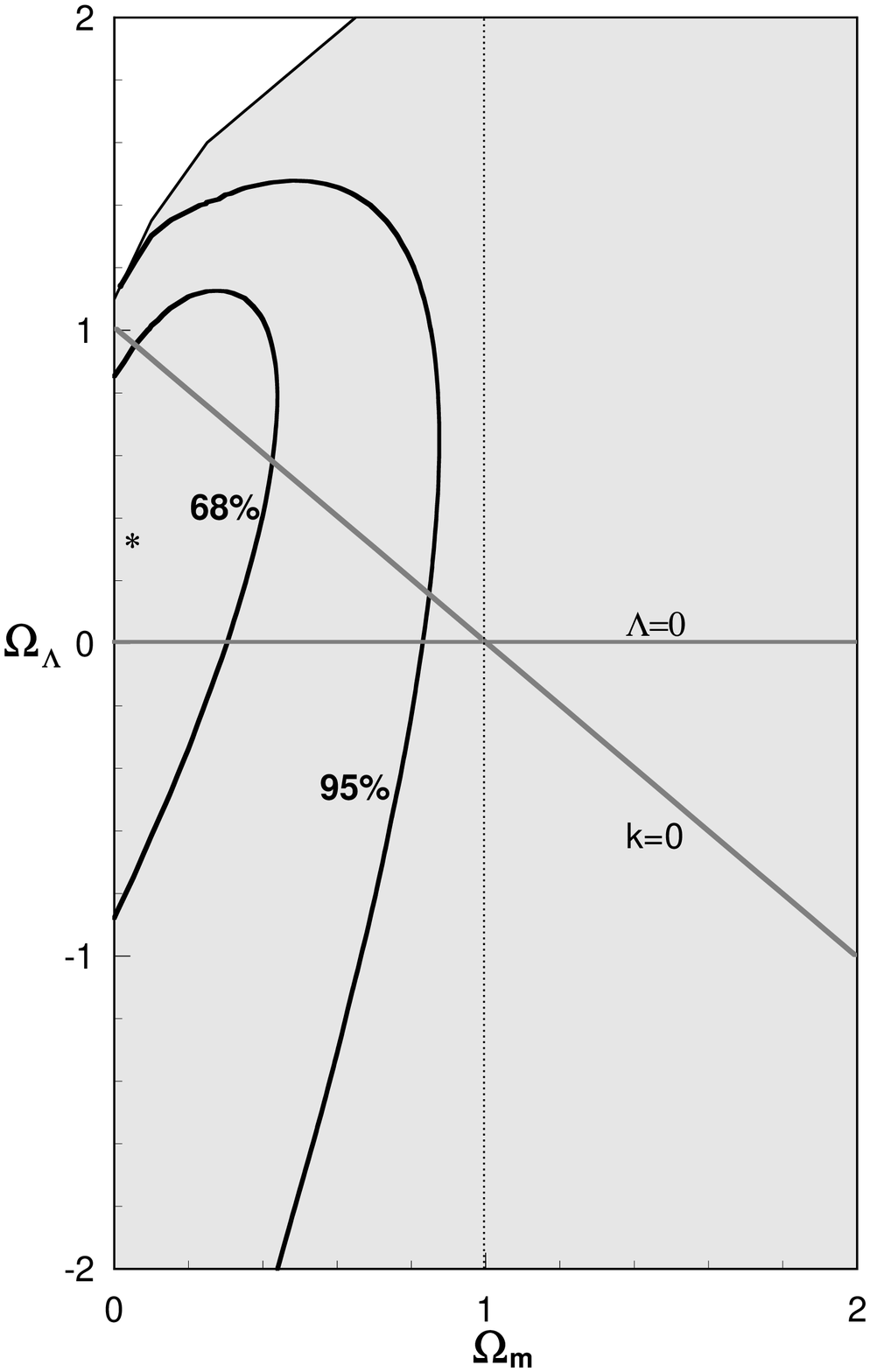,width=2.65in}
\caption{The data are compared with expectations in 
different cosmological models 
on the left hand panel.  Note how well the data track the curves
expected in a low density universe all the way from a redshift
of zero to a redshift of
about 2.  
The one-dimensional confidence contours on the 
parameters $\Omega_m$ and $\Omega_{\Lambda}$ are shown on 
the right hand panel; one and two sigma constraints on either parameter can 
be obtained by projecting onto the appropriate axis.  
A universe with $\Omega_m=1$ is
ruled out with 97.5 \% confidence.  Lines indicating
zero space curvature, zero cosmological constant, and
$\Omega_m=1$ are drawn 
on figure 1b.  The location of the minimum of the reduced
$\chi^2$ is indicated by a star.  }
\end{figure}

Clearly, $<D>$ for sources at a given redshift is proportional
to the angular size distance, luminosity distance, 
or coordinate distance $(a_or)$
to the sources.  For $\beta \simeq 2$, $D_* \propto
(a_or)^{-0.8}$, so the ratio $<D>/D_* \propto (a_or)^{1.8}$.
For the correct choice of cosmological parameters, the ratio $<D>/D_* =1$,
at all redshifts.  Thus, one way to compare the data with the 
cosmological 
models is to compare $(<D>/D_*)(a_or)^{-1.8}$ with the
curves $(a_or)^{-1.8}$ assuming different values for the 
cosmological parameters.  This is shown in figure 1a.  It is 
clear that a low value of $\Omega_m$ is favored, and $\Omega_m=1$
does not fit the data well at all.  In fact, it is shown
in figure 1b that $\Omega_m=1$ is ruled out with 97.5 \% 
confidence.  

Allowing for error bars on all quantities,  
the best fit values of the cosmological parameters 
are:  $\Omega_m = 0.2^{+0.3}_{-0.2}$ assuming zero space
curvature ($k=0$, $\Omega_{\Lambda} = 1 - \Omega_m$), 
and $\Omega_m = -0.1^{+0.5}_{-0.4}$ assuming zero cosmological
constant ($\Lambda=0$, $\Omega_k = 1-\Omega_m$).  
These values can be obtained by projecting the one-dimensional
fits, shown in figure 1b,  onto their respective axes.  Allowing 
$\Omega_{\Lambda}$
and $\Omega_k$ to vary simultaneously, the 
confidence levels for each parameter are shown on figure 1b
in such a way that the one and two sigma constraints on each
can be obtained by projecting onto the appropriate axis.

\section{Discussion}

The constraints on cosmological parameters obtained here
arise primarily from the relative position of the data
points as a function of redshift, and are not sensitive
to any particular point, or to the normalization of the 
curves (which is left as a free parameter in the fits).  
For example, if the one low-redshift
point near z = 0 is excluded, the results do not change
(see Guerra \& Daly 1998).  It is clear from figure 1a that
the radio source model is working extremely well; the 
data fall right along the cosmological curves all the 
way from zero redshift to a redshift of about two.  It is
clear from figure 1b that, given this data, it is very unlikely
that $\Omega_m=1$.  

The results presented here are very similar to those obtained
by the supernovae groups.  Any potential problems, such as
dust extinction or evolution, are completely different for
the two methods.  The fact that they yield nearly identical
results suggests that both are correct.  In addition,
the low value of $\Omega_m$ obtained using this method
agrees with the low values indicated by local dynamical
tests, which further suggests that $\Omega_m$ is low, and
the universe will continue to expand forever.

We are in the process of obtaining radio information on 
67 additional radio galaxies with redshifts between zero 
and two.  With these additional sources we will be able
to constrain cosmological parameters to very high accuracy.  

\section*{Acknowledgments}
It is a pleasure to thank Rick Perley and Ruth Durrer for 
helpful conversations.  This work was supported in part
by the US National Science Foundation, the Independent College
Fund of New Jersey, and a grant from Mark M. Wheeler III.

\end{document}